# Generalized Short Circuit Ratio for Grid Strength Assessment in Inhomogeneous Multi-infeed LCC-HVDC Systems

Guanzhong Wang, Huanhai Xin, Di Wu, Zhiyi Li, Linbin Huang, Ping Ju

*Abstract*—Generalized short circuit ratio (gSCR) for gird strength assessment of multi-infeed high voltage direct current (MIDC) systems is a rigorous theoretical extension of traditional short circuit ratio (SCR), which allows the considerable experience of using SCR to be extended to MIDC systems. However, gSCR was originally derived based on the assumption of homogeneous MIDC systems, where all HVDC converters have an identical control configuration, which poses challenges to the applications of gSCR to inhomogeneous MIDC systems. To weaken this assumption, this letter applies modal perturbation theory to explore the possibility of applying gSCR in inhomogeneous MIDC systems. Results of numerical experiments show that, in inhomogeneous MIDC systems, the previously proposed gSCR can still be used without modification, but critical gSCR (CgSCR) needs to be redefined by considering the characteristics of HVDC converter control configurations. Accordingly, the difference between gSCR and redefined CgSCR can effectively quantify the pertinent ac grid strength in terms of static voltage stability margin. The performance of our proposed method is demonstrated in a triple-infeed inhomogeneous LCC-HVDC system.

## I. INTRODUCTION

The line commutated converter-based high voltage direct current (LCC-HVDC) technique has been increasingly applied in the electric power grid for the long-distance and high-capacity power transmission, which boosts the development of multi-infeed dc systems (MIDC) where multiple HVDC inverters are connected to a common receiving end within close proximity [1]. In MIDC systems, static voltage instability issues may arise when he reactive power required by HVDC converters for their commutation is too large to support the grid voltage [2][3].

The ac grid strength plays a fundamental role in static voltage stability. Moreover, a simple measure named short circuit ratio (SCR) has long been used to quantify the grid strength in single-infeed LCC-HVDC (SIDC) systems. To be specific, the stability margin can be estimated by solely calculating SCR and critical SCR (CSCR), with CSCR≈2 in various SIDC systems [4]. To assess the grid strength for MIDC systems, several SCR-based methods have been proposed by considering the interactions among HVDC inverters [5-8]. These methods can be divided into two categories: empirical indices (including the multi-infeed interactive short circuit ratio (MISCR) [5] and the multi-infeed short circuit ratio (MSCR) [6]) and theoretical indices (including generalized effective short-circuit ratio (GESCR) [7] and generalized short circuit ratio (gSCR) [8]). On the one hand, the advantage of empirical indices is that their calculation formulas are simple; but they are short of theoretical justification due to their empirical reasoning, when the critical value of these indices may vary in different power systems. On the other hand, theoretical indices, e.g., GESCR, were theoretically proposed based on characteristic analysis of the Jacobian matrix, but the calculation formula of GESCR is much more complicated because it depends detailed system operation data. Furthermore, the critical GESCR is fixed at 1, which is quite different from SCR. Hence, the considerable experience of using the SCR cannot be simply adopted to the application of GESCR.

Compared with the above indices, gSCR keeps a simple calculation formula with a fixed critical gSCR (CgSCR), i.e., CgSCR=CSCR≈2 in various MIDC systems, because it was proposed by the theoretical analysis of the relationship between SCR and static voltage stability in SIDC systems and extending the results to MIDC systems [8]. This allows the use of the gSCR to enjoy the same experience of using the SCR. Particularly, the stability margin of MIDC systems can be solely focused on the gSCR and CgSCR. However, gSCR was derived based on the assumption of homogeneous MIDC systems, where all HVDC converters have the identical control configuration, which limits its applications to more general cases.

This letter is to extend the application of gSCR to inhomogeneous MIDC systems for grid strength assessment via mode perturbation theory. It will show that the gSCR can still be valid without modification by approximately deriving the relationship between gSCR defined for the homogeneous MIDC systems and the singularity point of the Jacobian matrix, but the CgSCR needs to consider the equivalent characteristic of a weighted sum of HVDC converter control configurations.

## II. PROBLEM STATEMENT

### A. Static Voltage Stability Analysis for MIDC Systems

The linearized power flow equations at the converter side of a MIDC system that is controlled by constant current-constant extinction angle or constant power-constant extinction angle can be represented as follows [9],

$$[\Delta \mathbf{P}_d \quad \Delta \mathbf{P} \quad \Delta \mathbf{Q}]^T = \mathbf{J}_{MIDC}[\Delta \mathbf{I}_d \quad \Delta \boldsymbol{\delta} \quad \Delta \mathbf{U}/\mathbf{U}]^T \quad (1)$$

where $\Delta \mathbf{P}_d$, $\Delta \mathbf{P}$, and $\Delta \mathbf{Q}$ are the vectors representing the perturbations of the DC power, active and reactive power at each converter-side ac bus; $\Delta \mathbf{I}_d$, $\Delta \boldsymbol{\delta}$, and $\Delta \mathbf{U}/\mathbf{U}$ are the vectors representing the perturbations of the DC current, voltage angle, and ac voltage percentage at each converter-side ac bus; $\mathbf{J}_{MIDC}$ is the Jacobian matrix.

This work was jointly supported by the National Natural Science Foundation of China under Grant 52007163 and China Postdoctoral Science Foundation funded project under Grant 2020M671718.

G. Wang (Email: eewangguanzhong@zju.edu.cn), H. Xin (Email: xinhh@zju.edu.cn; eexinhh@gmail.com), Z. Li and L. Huang (Email: huanglb@zju.edu.cn) are with Department of Electrical Engineering, Zhejiang University, Hangzhou 310027, China.

D. Wu is with the Department of Electrical and Computer Engineering, North Dakota State University, Fargo 58102, USA. (Email: di.wu.3@ndsu.edu).

It is known that the boundary condition for the static voltage stability in MIDC systems can be represented that the determinant of $J_{MIDC}$ is equal to zero (i.e., the saddle-node bifurcation).

$$\det(\mathbf{J}_{MIDC}) = 0 \tag{2}$$

This boundary condition in (2) can be simplified under the rated operating condition (i.e., $U_i = U_N = 1.0$ p.u. and $P_i = P_{Ni}$, $i=1,\ldots,n$) [8].

$$\det(\mathbf{J}_{MIDC}) = \det(\mathbf{J}_{sys}) = 0 \tag{3}$$

where $\mathbf{J}_{sys} = \text{diag}(T_i) + \mathbf{J}_{eq}^{-1} - \mathbf{J}_{eq}$ with $\mathbf{J}_{eq} = -diag^{-1}(P_{Ni})\mathbf{B}$; $P_{Ni}$ is the rated power injection into the ac grid from the $i^{th}$ converter; $T_i = 2cK(c)/[1-1/(\cos\gamma/c-1)] + 2\omega B_c U^2$, where $K(c)$ is a function of $c$ and $c = XI_d/(\sqrt{2}KU)$; $I_d$ is the DC current; $\gamma$ is the extinction angle; $K$ is the ratio of transformer; $U$ is the voltage magnitude; $X$ is the commutation reactance; $B_c$ is the reactive power compensation capacitor; $\omega$ is the angular velocity; $\mathbf{B}$ is the node susceptance matrix; more details can be found in [8].

*B. Challenge for Grid Strength Assessment Based on gSCR*

For a homogeneous MIDC system, the converters of all HVDC ties have the same control configuration. Thus, the parameter $T_i$ in (3) is an identical constant (i.e., $T=T_1=\ldots=T_i=\ldots=T_n$). $\mathbf{J}_{sys}$ can be rewritten as

$$\mathbf{J}_{sys0} = T \times I_n + \mathbf{J}_{eq}^{-1} - \mathbf{J}_{eq} \tag{4}$$

where $I_n$ is a $n$ by $n$ identity matrix.

By using mode decomposition technique for (4), the boundary condition in (3) can be further represented as [8]

$$\det(\mathbf{J}_{sys0}) = \prod_{i=1,2,\ldots,n}(T + \lambda_i^{-1} - \lambda_i) = 0 \tag{5}$$

where $T + \lambda_i^{-1} - \lambda_i$ and $\lambda_i$, (in the order of $0 < \lambda_1 \leq \cdots \lambda_i \leq \cdots \leq \lambda_n$) are eigenvalues of $\mathbf{J}_{sys0}$ and $\mathbf{J}_{eq}$, respectively.

Equation (5) is the product of eigenvalues of $\mathbf{J}_{sys0}$ and every eigenvalue of $\mathbf{J}_{sys0}$ can represent an equivalent SIDC systems for static voltage stability analysis [8]. Since the MIDC system stability mainly depends on the minimum eigenvalue of $\mathbf{J}_{sys0}$ or the equivalent SIDC system with $\lambda_1$, the boundary condition in (5) can be simplified as,

$$T + \lambda_1^{-1} - \lambda_1 = 0 \tag{6}$$

Based on (6), $\lambda_1$ is defined as gSCR such that the voltage stability margin of MIDC systems is quantified by the minimum eigenvalue of $\mathbf{J}_{eq}$ (a weighted node susceptance matrix), which significantly reduces the burden of voltage stability analysis with calculating the determinant of $J_{MIDC}$. Additionally, CgSCR is defined as the critical value of gSCR corresponding to the boundary condition in (6) and is represented by (7) below. In [4], it was found that CgSCR is approximately equal to 2 (the same value as CSCR in SIDC systems), which overcomes the bottleneck of ambiguity of critical values in the applications of SCR-based methods for MIDC systems [5][6].

$$\text{CgSCR} = T/2 + \sqrt{T^2/4 + 1} \tag{7}$$

where CgSCR is the positive root of the equation (6) with a single $\lambda_1$ variable.

It is noticed that gSCR can be analytically derived based on the assumption that each $T_i$ in (3) is equal in homogeneous MIDC systems. However, this assumption is not true in inhomogeneous MIDC systems, which limits the application of gSCR to inhomogeneous MIDC systems.

## III. Grid Strength Assessment

SCR based methods can be used to evaluate the stability margin of MIDC systems by focusing on the grid characteristics, i.e., network structure and parameters. For example, Section II introduced the concept of gSCR to quantitatively analyze the stability of homogeneous MIDC systems, where gSCR is the eigenvalue of the weighted node susceptance matrix $\mathbf{J}_{eq}$. However, in practice inhomogeneous MIDC systems (i.e., $T_1 \neq \ldots \neq T_i \neq \ldots \neq T_n$) also need to be investigated and the method in Section II is not applicable in such scenarios. To address this issue, the mode perturbation theory in [10] is employed to derive the relation between the stability of MIDC systems (reflected by the minimum eigenvalue of $\mathbf{J}_{sys}$) and the gSCR.

The following lemma provides the mathematical foundation for our proposed method.

***Lemma 1*** (Theorem 2.3 at page 183 in [10]). Let $\lambda$ be a simple eigenvalue of the matrix $A$, with right and left eigenvectors $x$ and $y$, and let $A+E$ be a perturbation of $A$. Then there is a unique eigenvalue $\tilde{\lambda}$ of $A+E$ such that

$$\tilde{\lambda} = \frac{y^T(\mathbf{A} + \mathbf{E})x}{y^T x} + O(\|E\|^2) \tag{8}$$

where the $O(\|E\|^2)$ is the second order small quality of $E$.

***Remark 1***: Let $\delta > 0$, $Y^T AX$, and $\varepsilon$ be the distance between $\lambda$ and the other eigenvalues of $A$, the Jordan canonical form of $A$, and the upper bound of $\|\mathbf{Y}\|\|\mathbf{E}\|\|\mathbf{X}\|$, respectively. If $E$ is so small that $16n\varepsilon^2/\delta^2 < 1$, then $\tilde{\lambda}$ is located uniquely in a Gerschgorin disk centered at $y^T(\mathbf{A}+\mathbf{E})x/(y^T x)$ with the radius bounded by $4n\varepsilon^2/\delta$ (seen in the proof of Theorem 2.3 [10]).

The minimum eigenvalue of $\mathbf{J}_{sys}$ for inhomogeneous systems can be derived by perturbing the minimum eigenvalue of $\mathbf{J}_{sys0}$ for the homogeneous systems based on lemma 1, which is summed as the following theorem.

***Theorem 1***: (a) The minimum eigenvalue of $\mathbf{J}_{sys}$ for inhomogeneous systems can be approximated as

$$\lambda_{\min}(\mathbf{J}_{sys}) = \mu_1^T\left[\text{diag}(T_i) + \mathbf{J}_{eq}^{-1} - \mathbf{J}_{eq}\right]v_1 = \sum_{j=1}^{n}\mu_{1,j}v_{1,j}T_j + \lambda_1 - \lambda_1^{-1} \tag{9}$$

and (b) the boundary condition $\det(\mathbf{J}_{sys}) = 0$ can be simplified as:

$$\lambda_{\min}(\mathbf{J}_{sys}) = \sum_{j=1}^{n}\mu_{1,j}v_{1,j}T_j + \lambda_1 - \lambda_1^{-1} = 0 \tag{10}$$

where $\mu_{1,j}$ and $v_{1,j}$ are the $j^{th}$ element of the left and right eigenvectors $\mu_1$ and $v_1$ of $\lambda_1$, respectively; $\sum_{j=1}^{n}\mu_{1,j}v_{1,j} = 1$ and $\mu_{1,j}v_{1,j} > 0$ [8].

***Proof***: $\text{diag}(T_i) + \mathbf{J}_{eq}^{-1} - \mathbf{J}_{eq}$ can be considered to be the perturbation of $\mathbf{J}_{sys0}$ whose eigenvectors are the same as those of $\mathbf{J}_{eq}$, so it follows from lemma 1 that its minimum eigenvalue can be approximated by $\mu_1^T\left[\text{diag}(T_i) + \mathbf{J}_{eq}^{-1} - \mathbf{J}_{eq}\right]v_1$, i.e., (a) is satisfied. Moreover, as the determinant of a matrix is equal to the product of its eigenvalues, the condition (b) is also satisfied. This concludes the proof.

***Remark 2***: The distance between converter control parameters $T_i$'s is generally smaller compared to the distance between decoupled ac grid structure parameter $\lambda_i$ in prevalent MIDC systems [1], which means the corresponding $\varepsilon$ and $\delta$ in Theorem 1 satisfy the condition $16n\varepsilon^2/\delta^2 < 1$ in Remark 1 and the approximation error of Theorem 1 is bounded by $4n\varepsilon^2/\delta \approx 0$. Moreover, if $16n\varepsilon^2/\delta^2 < 1$ is unsatisfied, (9) is still a good approximation in inhomogeneous systems, which can be illustrated by the cases in section IV where $\varepsilon \approx 0.2\delta$ and $16n\varepsilon^2/\delta^2 \approx 1.92$.

Equation (10) shows that the boundary condition for both homogeneous and inhomogeneous MIDC systems in (3) can be unified into equation (9) (i.e., replacing $T_i$ by $T$ in (9) yields (6)). Therefore, if gSCR= $\lambda_1$ and a modified CgSCR$^*$ in (11) are redefined for inhomogeneous systems, the voltage is stable if gSCR>CgSCR$^*$ and the voltage stability boundary can be approximated by the curve of gSCR=CgSCR*.

Similar to (7) for the homogeneous system, it follows from (10) that the CgSCR$^*$ for the inhomogeneous MIDC system can be defined as,

$$\text{CgSCR}^* = T_*/2 + \sqrt{T_*^2/4 + 1} \qquad (11)$$

where CgSCR$^*$ is the positive root of the equation (10) with a single $\lambda_1$ variable, and $T_* = \sum_{j=1}^{n} \mu_{1,j} \nu_{1,j} T_j$ is a weighted sum of $T_i$ of all HVDC converters in the MIDC systems.

It should be noticed that $T_*$ is in essence an equivalent HVDC control parameter in the corresponding SIDC system whose CSCR=CgSCR$^*$ and the extreme value of $T_*$ is determined by the existing HVDC control parameter $T_i$ in the MIDC system.

To implement the proposed method for stability studies of practical systems, the procedure of evaluating the system stability margin is shown in Fig. 1.

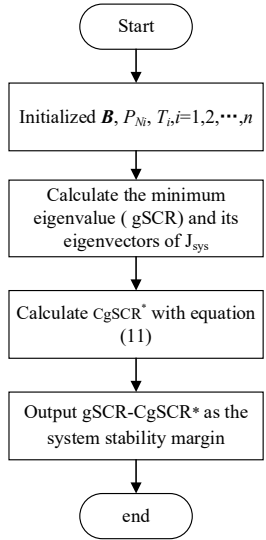

Fig. 1 Flowchart of evaluating the system stability margin using the proposed method

It should be noted that both the small-signal stability and the static voltage stability is the eigenvalue-based problem in power systems, the gSCR can be used in the two kind of problems [11]. So the idea for handling inhomogeneous power electronic devices can be used as well. For example, reference [12] shows that the idea can not only be extended for analyzing the small-signal stability of power systems with various renewable power generations such as PMSG and DFIG wind turbines, etc., but also it can be extended to improve the stability margin [13] by incorporating inhomogeneous devices which are deliberated designed.

## IV. NUMERICAL STUDIES

In this section, the effectiveness of gSCR and CgSCR$^*$ in (11) for grid strength assessment of inhomogeneous MIDC systems is demonstrated in an inhomogeneous triple-infeed HVDC system. The benchmark model proposed by CIGRE in 1991 [5] is applied here and the corresponding control configuration $T$=1.5. To highlight the inhomogeneity, by changing the commutation reactance, power-factor angle and transformer ratio of the benchmark model, three HVDC inverters that have different control parameters $T_i$ (e.g., $T_1$=1.24, $T_2$=1.5, $T_3$=1.75) are constructed. In addition, in the triple-infeed system [8], the Thevenin equivalent reactance is set as $z_1 = 1/1.5\,\text{p.u.}$, $z_2 = z_3 = 1/3\,\text{p.u.}$, $z_{12} = z_{13} = z_{23} = 1/1.5\,\text{p.u.}$.

Choose to verify the applicability of gSCR and CgSCR$^*$ to assess grid strength in terms of static voltage stability margin first. When increasing $P_{N2}$ and keep $P_{N1}$ and $P_{N3}$ constant, the gSCR and CgSCR$^*$ are evaluated. The changing results of gSCR and CgSCR$^*$ with $P_{N2}$ are shown in Fig. 1. It can be seen from Fig. 2 that gSCR decreases and CgSCR* tends to be constant as $P_{N2}$ increases. Thus, the static voltage stability margin quantified by the distance between gSCR and CgSCR$^*$ decreases as $P_{N2}$ increases. When $P_{N2}$ is increased to $P_{dmax}$ such that the determinant of $J_{MIDC}$ in (2) is equal to zero, gSCR coincides with CgSCR$^*$, which indicates that the static voltage stability limit occurs and thus stability margin is equal to zero.

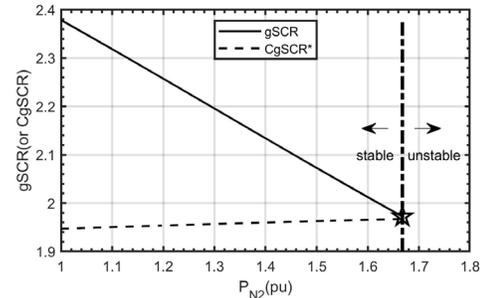

Fig. 2 Trajectories of gSCR and CgSCR with power PN2 for the triple-infeed system

Curves with different gSCR values (2, 2.1 and CgSCR$^*$, respectively) are all shown in Fig. 3, where the circles denote static voltage stability boundary ($J_{MIDC}$ in (1) is singular). To draw the curves, the rated power injections $P_{N1}$, $P_{N2}$, and $P_{N3}$ from those three HVDC ties are set up as follows: $P_{N3}$ maintains 1 p.u., $P_{N2}$ varies from 1 p.u. to 1.4 p.u., and $P_{N1}$ is changing in order to make $J_{MIDC}$ singular or gSCR coincide with different values. It can be seen from Fig. 3 that the static voltage stability boundary and the curve with gSCR=CgSCR$^*$ are very close. Especially, the largest relative error between the points on the static voltage stability boundary and those on the curve with gSCR=CgSCR$^*$ is only 0.41% by fixing $P_{N1}$ and comparing $P_{N2}$ in the curves. In conclusion, the voltage stability boundary can be well approximated by the curve of gSCR=CgSCR$^*$. Moreover, the larger value of gSCR denotes the larger stability margin because the curve with a larger gSCR is closer to the origin point than those with a smaller gSCR.

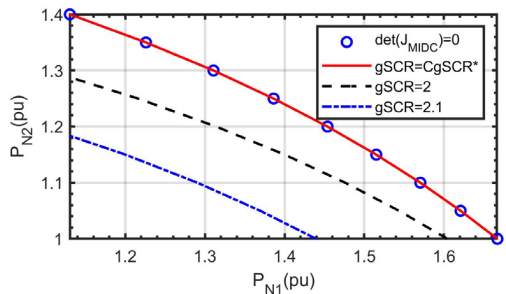

Fig. 3 Trajectories of power PN2 with PN1 responding to different gSCR's and a singular Jacobian matrix

The relative error between CgSCR* and gSCR at the stability boundary is further analyzed when the inhomogeneity level in HVDC inverters changes in the system. The inhomogeneous level is quantified by the standard deviation of control parameters $T_i$ ($i$=1, 2, 3) of those three HVDC inverters. Table 1 presents the largest percentage error as well as the standard deviation of $T_i$, when $T_1$ and $T_3$ change but $T_2$ keeps constant. It is observed from this table that the approximation error of stability boundary by using CgSCR* is insensitive to changes in control parameters, since the largest percentage error is small even when $T_1$=1.0439 and $T_3$=1.9245 are significantly different from the benchmark model with $T$=1.5 ($\varepsilon \approx 0.2\delta$).

TABLE 1
ERROR ANALYSIS FOR THE TRIPPLE-INFEED SYSTEM

| $T_1$ | $T_3$ | Standard deviation of $T_i$'s | Error level |
|---|---|---|---|
| 1.2444 | 1.7455 | 0.2505 | 0.33% |
| 1.1786 | 1.8056 | 0.3135 | 0.52% |
| 1.1118 | 1.8652 | 0.3768 | 0.75% |
| 1.0439 | 1.9245 | 0.44 | 1.01% |

III. CONCLUSION

The modal perturbation theory was used to extend the application of the gSCR previously defined for homogeneous MIDC systems to inhomogeneous MIDC systems. It was demonstrated that the difference between gSCR and a modified CgSCR* can effectively assess grid strength of inhomogeneous HVDC in terms of static voltage stability margin. Moreover, the proposed CgSCR* is a promising way to estimate the static voltage stability limit under various HVDC control parameters, which is our future work.

REFERENCES

[1] E. Rahimi, "Voltage interactions and commutation failure phenomena in multi-infeed HVDC systems," Ph.D. dissertation, Dept. Elect. Comput. Eng., Univ. Manitoba, Winnipeg, MB, Canada, 2011.
[2] Aik D L H, Andersson G. Analysis of voltage and power interactions in multi-infeed HVDC systems[J]. IEEE Transactions on Power Delivery, 2013, 28(2): 816-824.
[3] H. Xiao and Y. Li, "Multi-Infeed Voltage Interaction Factor: A Unified Measure of Inter-Inverter Interactions in Hybrid Multi-Infeed HVDC Systems," *IEEE Transactions on Power Delivery*, vol. 35, no. 4, pp. 2040-2048, Aug. 2020.
[4] P. C. S. Krishayya, R. Adapa, and M. Holm, "IEEE guide for planning DC links terminating at AC locations having low short-circuit capacities, part I: AC/DC system interaction phenomena," *CIGRE*, France, 1997.
[5] B. Davies, A. Williamson, A. M. Gole, B. EK, B. Long, B. Burton, D. Kell, D. Brandt, D. Lee, E. Rahimi, G. Andersson, H. Chao, I. T. Fernando, K. L. Kent, K. Sobrink, M. Haeusler, N. Dhaliwal, N. Shore, P. Fischer, and S. Filizadeh, "System with multiple DC infeed," in *CIGRE Working Group B4.41*, Dec. 2008.
[6] P. F. de Toledo, B. Bergdahl, and G. Apslund, "Multiple infeed short circuit ratio-Aspects related to multiple HVDC into one AC network," *IEEE/PES Transmission and Distribution Conference and Exhibition: Asia and Pacific.*, pp. 1-6, 2005.
[7] H. Xiao, Y. Li, D. Shi, J. Chen and X. Duan, "Evaluation of Strength Measure for Static Voltage Stability Analysis of Hybrid Multi-Infeed DC Systems," *IEEE Transactions on Power Delivery*, vol. 34, no. 3, pp. 879-890, June 2019.
[8] F. Zhang, H. Xin, D. Wu, Z. Wang, and D. Gan, "Assessing grid strength of multi-infeed lcc-hvdc systems using generalized short circuit ratio," *IEEE Trans. Power Syst.*, vol. 34, no. 1, pp. 467–480, Jan. 2019.
[9] D. L. H. Aik and G. Andersson, "Power stability analysis of multi-infeed HVDC systems," *IEEE Transactions on Power Delivery*, vol. 13, no. 3, pp. 923-931, July 1998.
[10] Stewart G W. Matrix perturbation theory[J]. 1990.
[11] Wei Dong, Huanhai Xin, Di Wu, Small Signal Stability Analysis of Multi-Infeed Power Electronic Systems Based on Grid Strength Assessment, IEEE Trans. on Power Systems, vol. 34, no. 2, pp. 1393-1403, 2019.
[12] H. Xin, D. Gan, P. Ju, Generalized Short Circuit Ratio of Power Systems with Multiple Power Electronic Devices: Analysis for Various Renewable Power Generations, Proceeding of the CSEE, 2020-08-06 (in Chinese).
[13] Chaoran Yang, Linbin Huang, Huanhai Xin, et al., Placing Grid-Forming Converters to Enhance Small Signal Stability of PLL-Integrated Power Systems, IEEE Trans. on Power Systems, early access, 2020.